\documentclass[titlepage,12pt]{utarticle}
\usepackage{amsmath,amssymb,amsthm,amscd,cite,epsfig}
\usepackage{latexsym,amsmath,amssymb,graphicx}
\usepackage[all]{xy}
\usepackage{epsf}
\xyoption{v2}
\xyoption{2cell}
\xyoption{dvips}
\CompileMatrices
\LaTeXdiagrams
\UseAllTwocells
\newdimen\tableauside\tableauside=1.0ex
\newdimen\tableaurule\tableaurule=0.4pt
\newdimen\tableaustep
\def\phantomhrule#1{\hbox{\vbox to0pt{\hrule height\tableaurule width#1\vss}}}
\def\phantomvrule#1{\vbox{\hbox to0pt{\vrule width\tableaurule height#1\hss}}}
\def\sqr{\vbox{%
  \phantomhrule\tableaustep
  \hbox{\phantomvrule\tableaustep\kern\tableaustep\phantomvrule\tableaustep}%
  \hbox{\vbox{\phantomhrule\tableauside}\kern-\tableaurule}}}
\def\squares#1{\hbox{\count0=#1\noindent\loop\sqr
  \advance\count0 by-1 \ifnum\count0>0\repeat}}
\def\tableau#1{\vcenter{\offinterlineskip
  \tableaustep=\tableauside\advance\tableaustep by-\tableaurule
  \kern\normallineskip\hbox
    {\kern\normallineskip\vbox
      {\gettableau#1 0 }%
     \kern\normallineskip\kern\tableaurule}%
  \kern\normallineskip\kern\tableaurule}}
\def\gettableau#1 {\ifnum#1=0\let\next=\null\else
  \squares{#1}\let\next=\gettableau\fi\next}
\numberwithin{equation}{section}
\xyoption{v2}
\xyoption{2cell}
\xyoption{dvips}
\CompileMatrices
\LaTeXdiagrams
\UseAllTwocells
\newdimen\tableauside\tableauside=1.0ex
\newdimen\tableaurule\tableaurule=0.4pt
\newdimen\tableaustep
\def\phantomhrule#1{\hbox{\vbox to0pt{\hrule height\tableaurule width#1\vss}}}
\def\phantomvrule#1{\vbox{\hbox to0pt{\vrule width\tableaurule height#1\hss}}}
\def\sqr{\vbox{%
  \phantomhrule\tableaustep
  \hbox{\phantomvrule\tableaustep\kern\tableaustep\phantomvrule\tableaustep}%
  \hbox{\vbox{\phantomhrule\tableauside}\kern-\tableaurule}}}
\def\squares#1{\hbox{\count0=#1\noindent\loop\sqr
  \advance\count0 by-1 \ifnum\count0>0\repeat}}
\def\tableau#1{\vcenter{\offinterlineskip
  \tableaustep=\tableauside\advance\tableaustep by-\tableaurule
  \kern\normallineskip\hbox
    {\kern\normallineskip\vbox
      {\gettableau#1 0 }%
     \kern\normallineskip\kern\tableaurule}%
  \kern\normallineskip\kern\tableaurule}}
\def\gettableau#1 {\ifnum#1=0\let\next=\null\else
  \squares{#1}\let\next=\gettableau\fi\next}
\newcommand{\be}{\begin{equation}}
\newcommand{\ee}{\end{equation}}
\newcommand{\refeq}[1]{(\ref{eq:#1})}
\newcommand{\IP}{\mathbb{P}}

\newcommand\IZ{\mathbb {Z}}
\newcommand{\IC}{\mathbb{C}}

\newcommand{\ba}{\begin{array}}
\newcommand{\ea}{\end{array}}

\newcommand{\lb}{\lambda}

\newcommand{\CK}{{\cal K}}
\newcommand{\kah}{{K\"ahler}}

\newcommand{\bal}{\begin{aligned}}
\newcommand{\eal}{\end{aligned}}

\newcommand{\re}{{\mathrm {Re}}}
\newcommand{\im}{{\mathrm {Im}}}
\newcommand{\IQ}{{\mathbb Q}}
\newcommand{\half}{{1\over 2}}
\DeclareFontFamily{U}{rsf}{}
\DeclareFontShape{U}{rsf}{m}{n}{
  <5> <6> rsfs5 <7> <8> <9> rsfs7 <10-> rsfs10}{}
\DeclareMathAlphabet\Scr{U}{rsf}{m}{n}

\begin{document}
\preprint{
    {\tt hep-th/0701104}\\
    RUNHETC-07-34\\
    SU-ITP-06/29\\
    SLAC-PUB-12180\\
}
\title{Metastable Quivers in String Compactifications}
\author{Duiliu-Emanuel Diaconescu,$^{\flat}$\footnote{{\tt 
duiliu@physics.rutgers.edu}}~ Ron Donagi,$^{\natural}$\footnote{{\tt 
donagi@math.upenn.edu}}~
and Bogdan Florea$^{\sharp}$\footnote{{\tt bflorea@slac.stanford.edu}}}
\oneaddress{
     {\centerline  {$^{\flat}$\it Department of Physics and Astronomy, 
     Rutgers University,}}
     \smallskip
     {\centerline {\it Piscataway, NJ 08854-0849, USA}}
      \smallskip
{\centerline {$^{\natural}$\it Department of Mathematics, University of 
Pennsylvania,}}
\smallskip 
{\centerline {\it David Rittenhouse Lab., 209 South 33rd St., Philadelphia
PA 19104-6395}}
      {\centerline {$^{\sharp}$ \it Department of Physics and SLAC, Stanford University,}}
      \smallskip
      {\centerline {\it Palo Alto, CA 94305, USA}}}
\date{January 2007}

\Abstract{We propose a scenario for dynamical supersymmetry breaking 
in string compactifications based on geometric engineering of quiver 
gauge theories. In particular we show that the runaway behavior of 
fractional branes at del Pezzo singularities can be stabilized by 
a flux superpotential in compact models. Our construction relies 
on homological mirror symmetry for orientifolds.}

\maketitle

\section{Introduction}

It has been observed in
\cite{Berenstein:2005xa,Franco:2005zu,Bertolini:2005di}
that nonperturbative effects induce dynamical supersymmetry 
breaking in certain quiver gauge theories. These quiver gauge 
theories can be engineered in terms of fractional branes at 
Calabi-Yau threefold singularities. Therefore this setup 
is a natural candidate for a supersymmetry breaking mechanism 
in string theory. 

However it has been shown in 
\cite{Intriligator:2005aw} (see  also 
\cite{Franco:2005zu,Diaconescu:2005pc,Forcella:2006ab})  
that these models give rise to runaway behavior in the \kah\ moduli space 
when embedded in string compactifications. More precisely, in the absence 
of a moduli stabilization mechanism, the closed string \kah\ parameters 
-- which couple to brane world-volume actions as FI terms -- 
can take arbitrary values. Therefore the D-flatness conditions 
are automatically satisfied, and one does not obtain a metastable 
non-supersymmetric vacuum. This problem is present in the F-theory 
models developed in \cite{Diaconescu:2005pc}, in which case the 
\kah\ moduli should be stabilized by Euclidean D3-brane instanton effects
\cite{EW:Minst} by analogy with \cite{DDF,DDKF,KKLT,Lust:2005dy,Lust:2006zg,
Balasubramanian:2005zx}. In the present context 
such instantons develop extra zero modes as a result of their interaction 
with the fractional branes, which can in principle lead to cancellations 
in the effective superpotential \cite{Ganor:1998ai}. Therefore one cannot 
rule out the existence of runaway directions in the \kah\ moduli space 
without a more thorough analysis. Some progress in this direction has been 
recently made in \cite{Argurio:2006ew,Florea:2006si,Argurio:2006ny}
(see also \cite{Blumenhagen:2006xt} for a discussion of D-brane instanton 
effects in IIA models.) 
 
In this paper we propose an alternative embedding scenario of supersymmetry 
breaking quivers in string compactifications. The starting point of our 
construction is the observation that quiver gauge theories typically occur 
in nongeometric phases in the \kah\ moduli space. 
These are regions of the $N=2$ \kah\ moduli space where the 
quantum volumes of certain holomorphic cycles become of the 
order of the string scale.
As a result, the dynamics of IIB $N=1$ orientifold models is very 
hard to control in this regime, since the supergravity approximation 
is not valid. 

It has been long known that nongeometric phases in $N=2$ compactifications
become more tractable in the mirror description of the models.
Mirror symmetry identifies the \kah\ moduli space of a Calabi-Yau 
threefold $X$ with the complex structure moduli space of the mirror 
threefold $Y$. The nongeometric phases of $X$ are mapped to certain 
regions in the complex structure moduli space of $Y$. One can maintain 
at the same time the volume of the mirror threefold $Y$ large, 
obtaining therefore a large radius compactification, where the 
supergravity approximation is valid. This idea has been implemented  
in the context of $N=1$ string vacua in \cite{DGS:land}. 

Following the same strategy, in section two we propose a construction 
of supersymmetry breaking quivers in terms of D6-brane configurations 
in IIA Calabi-Yau orientifolds. This construction relies on homological 
mirror symmetry for orientifold models, but does not require 
an extension of mirror symmetry to nonzero flux. 
Note that a different dynamical supersymmetry breaking mechanism 
in IIA toroidal orientifolds has been proposed in \cite{Cvetic:2003yd}.

In addition to 
solving the problem of small quantum volumes, this scenario also 
offers a natural moduli stabilization superpotential induced by 
background fluxes. In particular we will show in section three that 
the potential runaway directions in the moduli space can be very efficiently 
stabilized by turning on IIA NS-NS flux. The IIA vacuum structure in 
the presence of fluxes has been previously investigated in 
\cite{Behrndt:2003ih, 
Dall'Agata:2003ir,Kachru:2004jr,DeWolfe:2004ns,
Derendinger:2004jn,Behrndt:2004km,Tsimpis:2004ab,Camara:2005pr,Camara:2005dc,
Villadoro:2005cu,Saueressig:2005es}.
Note that in this picture 
closed string moduli stabilization occurs at a much higher scale 
than the typical quiver gauge theory energy scale. The low energy 
effective dynamics of the system reduces to open string dynamics 
in a fixed closed string background, and is dominated by strong 
infrared effects in the Yang-Mills theory. A similar hierarchy of scales 
occurs in the construction of de Sitter vacua in string theory
proposed in \cite{KKLT}. 

In section four we construct concrete compact models 
satisfying the general conditions formulated in 
section three. We exhibit a landscape 
of vacua equipped with supersymmetry breaking quivers
in a specific example, which 
is very similar to the ones discussed in section 6 of 
\cite{Diaconescu:2005pc}. An interesting problem for future work
is whether the present construction of supersymmetry breaking vacua 
is compatible with the fractional brane realization of the MSSM
\cite{Verlinde:2005jr,Buican:2006sn}. In particular, it would be 
very interesting to find stabilized string vacua which can accommodate 
both supersymmetry breaking quivers and MSSM fractional brane 
configurations. This would be a concrete realization of the 
ideas outlined in \cite{Diaconescu:2005pc}.

{\it Acknowledgments.}
We owe special thanks to Shamit Kachru for
a careful reading of the manuscript and many valuable suggestions and
comments and to Robert Karp for very helpful discussions on 
del Pezzo fractional branes, periods and monodromy.
We would also like to thank Mike Douglas,
Chris Herzog, Tony  Pantev, 
David Shih and especially Scott Thomas for
very stimulating discussions.
D.-E. D. was partially supported by NSF grant
PHY-0555374-2006.
R.D. was partially supported by NSF grants DMS 0104354, DMS 0612992
and NSF Focused Research Grant DMS 0139799. 
The work of B. F. was supported in part by the NSF under grant
PHY-0244728 and DOE under contract DE-AC03-76SF00515.

\section{Supersymmetry Breaking and del Pezzo Surfaces in String 
Compactifications}

In this section we review the construction of supersymmetry breaking 
quivers in terms of fractional branes at del Pezzo singularities, 
and propose an embedding strategy in IIA string compactifications. 
Since the gauge theory dynamics has been thoroughly analyzed in 
\cite{Berenstein:2005xa,Franco:2005zu,Bertolini:2005di}, we will focus 
only on the relevant string theory aspects.  

\subsection{Fractional Branes and Supersymmetry Breaking Quivers} 

The first examples of supersymmetry breaking quivers 
\cite{Berenstein:2005xa,Franco:2005zu,Bertolini:2005di} 
were realized in terms of fractional D5-branes at a $dP_1$ singularity. 
Recall that the first del Pezzo surface $dP_1$ is a one point blow-up 
of the projective plane $\IP^2$. The Picard group of $dP_1$ is 
generated by the hyperplane class $h$ and the class of the exceptional 
divisor $e$. 
Under Calabi-Yau/Landau-Ginzburg correspondence, the fractional 
branes at a $dP_1$ singularity are in one to one correspondence 
with the following collection of bundles\footnote{Here we denote by 
${\overline E}$ the anti-brane 
of a D-brane with Chan-Paton bundle $E$. This notation is not 
quite rigorous from the point of view of the derived category, but it will 
suffice for our purposes.} on $dP_1$ (see for example 
\cite{Herzog:2003dj})
\be\label{eq:quiverA} 
E_1 = \CO_{dP_1}, \qquad E_2 = {\overline{\CO_{dP_1}(h-e)}}, 
\qquad E_3={\overline{\CO_{dP_1}(e)}}, \qquad 
E_4 = \CO_{dP_1}(h).
\ee
The quiver gauge theory is engineered by a brane 
configuration of the form 
\be\label{eq:quiverB} 
E_1^{\oplus (N+M)}\oplus  E_2^{\oplus (N+3M)} \oplus E_3^{\oplus N}  
\oplus E_4^{\oplus (N+2M)}  
\ee
where $N,M$ are positive integers. This gives rise to the quiver 
represented in fig. 1.  
\begin{figure}[ht]
\centerline{\epsfxsize=4.5cm \epsfbox{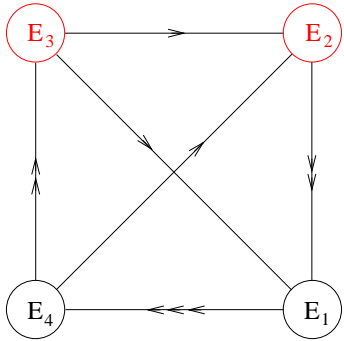}}
\caption{Quiver gauge theory associated to a $dP_1$ singularity.} 
\label{fig:quiver} 
\end{figure}
In terms of large radius charges, 
one can show that this configuration has 
$N$ units of D3-brane charge and $M$ units of  
D5-brane charge wrapped on a two-cycle $\Gamma$ orthogonal to the 
canonical class $K_{dP_1}$. The net D7-brane charge of the configuration 
\eqref{eq:quiverB} is zero. 


For the purpose of embedding in string compactifications 
it is important to note that the quiver theory in fig. 1 can be 
alternatively engineered using other  
$dP_k$ singularities with various values of $1\leq k\leq 8$. 
For example according to section 4.1 of 
\cite{Diaconescu:2005pc}, one can obtain the same quiver 
gauge theory from 
fractional branes at $dP_8$ or $dP_5$ singularities.  
Similar constructions can be carried out for $dP_6$ and $dP_7$ 
singularities as well. 
We will not need the details of these constructions in the following. 
The essential point for us is that in each case the quiver gauge theory 
will be engineered by a collection of fractional branes 
of the form 
\be\label{eq:quiverC} 
\oplus_{a=1}^s E_a^{\oplus N_a} 
\ee
where $E_1,\ldots,E_s$
are exceptional bundles on the del Pezzo surface, and 
$N_a$ are positive integers. The bundles $E_a$ are stable and have 
no infinitesimal deformations (they are usually called spherical 
objects in derived category language.) 

\subsection{Strategy For Embedding in String Compactifications} 

Let us now present our strategy for embedding the local 
del Pezzo models in string compactifications. 
As explained in the introduction, we will construct IIA 
orientifold models related by mirror symmetry to IIB 
compactifications with fractional branes at del Pezzo singularities. 
 
We start with a few preliminary remarks on del Pezzo singularities 
in IIB Calabi-Yau compactifications. Suppose we have a Calabi-Yau 
threefold $X$ and a contraction map
\[ 
c : X \to {\widehat X} 
\] 
which collapses a smooth del Pezzo surface $S$ to a singular point 
of ${\widehat X}$. Moreover, suppose that this degeneration occurs 
on a codimension $k$ wall in the \kah\ cone of $X$. 
This means that we have to tune $k$ \kah\ moduli of $X$ in 
order to reach the singularity i.e. the image of the restriction map 
\[ 
H^{1,1}(X) \to H^{1,1}(S) 
\]
has rank $k$. 
Note that in local models $X$ is isomorphic to the total space 
of the canonical bundle $K_S\to S$, and $k=h^{1,1}(S)$. 

As explained earlier in this section, the fractional branes 
associated to a del Pezzo singularity are related by analytic 
continuation to an exceptional collection
$E_1,\ldots,E_s$ on the collapsing del Pezzo surface.  
The quiver gauge theory 
occurs along a subspace of the \kah\ moduli space where the
$N=2$ central charges of the branes $E_1,\ldots,E_s$ are aligned. 
We will refer to this subspace as the quiver subspace, or the quiver locus,
in the following. A detailed analysis of this locus in a local 
del Pezzo model can be found in \cite{Aspinwall:2004vm}. We will 
perform a similar computation in a compact model in section four. 

Note that the central charges are aligned, but nonzero, along 
the quiver locus, therefore 
the underlying $N=2$ SCFT is not singular.
In other words there are no massless states or tensionless 
extended objects associated 
to the collapsing del Pezzo surface since the quantum volumes of the 
stable branes wrapping various cycles therein do not vanish. 
From this point of view the quiver subspace is analogous to a 
Landau-Ginzburg phase, as opposed to a conifold singularity. 
It follows however that world-sheet instanton corrections are 
important near the quiver locus of the moduli space, and 
the supergravity approximation is not valid. 
This can be a serious drawback in the context of $N=1$ 
compactifications, especially compactifications with background 
flux since the dynamics
is under control only in the large radius regime. 

This problem can be avoided employing the strategy developed in 
\cite{DGS:land}. Namely, using homological mirror symmetry, we map 
the IIB fractional brane configuration to a configuration of 
D6-branes wrapping  special lagrangian cycles in a IIA compactification 
on the mirror threefold $Y$. The dynamics can be kept under control 
by working near the large complex limit point in the complex structure 
moduli space of $X$, which corresponds to the large radius limit
in the \kah\ moduli space of $Y$. We stress that this statement relies 
only on the standard homological mirror symmetry conjecture, and 
eventually its extension to orientifold models. We do not need to invoke 
any form of mirror symmetry in the presence background flux, which would 
be a more delicate issue. As in \cite{DGS:land}, the type IIB description 
of the theory is mainly used here as a convenient mathematical tool which 
makes certain geometric aspects more transparent. Once we have established the 
existence of a suitable IIA D6-brane configuration in a large volume 
compactification, the IIB theory ceases to play any role. Flux dynamics 
and moduli stabilization will be discussed only in terms of the IIA 
model, in which the supergravity approximation is valid. In principle 
one could formulate the whole discussion in IIA variables from the 
beginning, but then it would be more difficult to establish the existence 
of a Calabi-Yau threefold $Y$ with the desired properties. 

Homological mirror symmetry predicts the existence of a collection 
$(M_1,V_1),\ldots,(M_s,V_s)$ of special lagrangian 
three-cycles\footnote{In principle one may wonder if the mirror 
{\bf A}-branes of the del Pezzo fractional branes may be stable 
coisotropic branes \cite{Kapustin:2001ij,Kapustin:2003kt} 
rather than special lagrangian 
three-cycles. The role of such objects in homological mirror 
symmetry and their phenomenological implications have been discussed 
in \cite{Kapustin:2001ij,Kapustin:2003kt,Grange:2004ah}
and \cite{Font:2006na} respectively. Although we cannot rigorously 
rule out this possibility, we would like to point out that it is 
rather unlikely for del Pezzo fractional branes. For local models 
it was found for example in \cite{Hanany:2001py} that del Pezzo 
fractional branes are indeed related to special lagrangian 
three-cycles by mirror symmetry. Since local models are obtained 
in fact from compact ones by taking a certain (analytic) limit 
\cite{Klemm:1996hh,Lerche:1996ni,Chiang:1999tz}, the same conclusion 
will conjecturally hold in compact models as well. We thank S. Kachru for 
a discussion on these issues.}
$M_1,\ldots,M_s$
equipped with flat unitary bundles $V_a$ corresponding
to the exceptional objects $E_1,\ldots,E_s$. Since the later are 
spherical objects in the derived category of $X$, $M_1,\ldots,M_s$ 
must be lagrangian three-spheres in $Y$. In particular the bundles 
$V_a$ will be topologically trivial. 
Moreover, since the fractional branes are indecomposable objects 
in the derived category of $X$, the bundles $V_a$ must have rank $1$. 
Therefore each object $E_a$ corresponds to a single D6-brane 
wrapping a special lagrangian three-sphere in $Y$. 
The central charge of such a brane is 
\be\label{eq:chcharge} 
Z_a = \int_{M_a} \Omega_Y.
\ee 
As explained above, the quiver locus is a special subspace 
in the complex structure 
moduli space of $Y$ where all $Z_a$ are aligned. 

Since $X,Y$ are compact threefolds, the model should 
also include an orientifold projection in order for tadpole 
cancellation conditions to be satisfied. 
We will impose a IIB orientifold 
projection of the form 
\be\label{eq:orientifA} 
(-1)^{F_L}\Omega \sigma 
\ee
where $\sigma :X \to X$ is a holomorphic 
involution so that the fixed point set $X^\sigma$ consists 
of zero and two complex dimensional components. 
This defines a mixed O3/O7 IIB orientifold model. 
We will also require several additional conditions to be satisfied. 

\begin{itemize} 
\item[$(i)$] The threefold $X$ contains a pair $(S,S')$ of disjoint 
del Pezzo surfaces which do not intersect the fixed locus 
$X^\sigma$ so that $\sigma : S \to S'$ is an isomorphism. 
We will refer to such a pair as conjugated del Pezzo surfaces.
Although most considerations in this paper hold for arbitrary 
del Pezzo surfaces $S,S'$ we will restrict ourselves to models 
in which $S,S' \simeq dP_n$, $n=6,7,8$. This restriction will facilitate 
the analysis of the critical locus of the flux superpotential in the next 
section. 

\item[$(ii)$] The holomorphic involution should be compatible with 
the large complex structure limit in the complex structure moduli 
space of $X$. This means we should be able to find a family of 
threefolds $X$ equipped with involutions $\sigma$ passing arbitrarily 
close to the large complex structure limit point. 
This condition is needed in order 
for the mirror IIA theory to be a large radius compactification. 
\end{itemize} 

Assuming these conditions satisfied, orientifold mirror symmetry 
\cite{Acharya:2002ag,Brunner:2003zm,Brunner:2004zd,Grimm:2004ua}
implies 
that a pair $(X,\sigma)$ is related to a pair $(Y,\eta)$ where 
$\eta:Y \to Y$ is an anti-holomorphic involution of $Y$. 
The D6-brane configuration $(M_a,V_a)$ is mapped to a conjugate
configuration $(M_a',V_a')$ so that 
$\eta:M_a\to M_a'$ is an isomorphism and $\eta^*(V_a') = V_a$. 
If the complex structure of $X$ is near the large complex structure point, 
$Y$ will 
determine a large volume IIA compactification. This is the desired 
embedding of del Pezzo quiver theories in string compactifications. 
Next we address the issue of moduli stabilization. 

\section{IIA moduli stabilization} 

In this section we 
analyze the effective four dimensional supergravity theory 
of the IIA compactification using the formalism of \cite{Grimm:2004ua}. 
Our goal is to argue for the existence of stabilized supersymmetric 
IIA vacua so that the complex structure moduli of $Y$ are fixed on the
quiver locus. 

The \kah\ moduli of $Y$ can be stabilized by turning on generic RR flux
$F=F_0+F_2+F_4+F_6$ on $X$. This gives rise to a superpotential of the 
form \cite{Gukov:1999ya,Gukov:2002iq}
\be\label{eq:superpotA}
W_K = \int_Y F \wedge e^{-J_Y}
\ee
where $J_Y$ is the \kah\ class of $Y$. For generic fluxes, this
superpotential depends on all \kah\ moduli, therefore we expect all 
flat directions in the \kah\ moduli space to be lifted. 

Complex moduli stabilization is a more complicated problem in the 
present  situation. In order to fix the complex structure moduli of $Y$
we have to turn on the IIA NS-NS flux $H$ \cite{Behrndt:2003ih, 
Dall'Agata:2003ir,Kachru:2004jr, Grimm:2004ua,DeWolfe:2004ns,
Derendinger:2004jn,Behrndt:2004km,Tsimpis:2004ab,Camara:2005pr,Camara:2005dc,
Villadoro:2005cu,Saueressig:2005es}. In the presence of D6-branes 
the flux is constrained however by the Freed-Witten anomaly cancellation 
condition \cite{Freed:1999vc}, which states that 
\be\label{eq:FWanom} 
\int_M H =0 
\ee
for any three-cycle $M$ supporting a D6-brane. This can in principle 
hinder moduli stabilization by disallowing a sufficiently generic 
$H$ flux. This was previously observed in the context of IIA 
flux vacua in \cite{Camara:2005dc}, and a supergravity interpretation 
of this constraint has been developed in \cite{Villadoro:2005yq}.
In particular, we have to enforce condition \eqref{eq:FWanom}
for all cycles $M\subset Y$ supporting D6-branes related by mirror 
symmetry to fractional branes. 
  
In order to address this problem
we need a detailed description of the effective supergravity action 
of IIA orientifolds, which has been given in \cite{Grimm:2004ua}. 
Following \cite{Grimm:2004ua}, let us 
choose a symplectic basis of three-cycles 
$(\alpha_i,\alpha_\lambda; \beta^i, \beta^\lambda)$ on $Y$ 
so that $(\alpha_i,\beta^\lambda)$ 
span $H^3_+(Y)$ and $(\alpha_\lambda, \beta^i)$ span $H^3_-(Y)$.
The indices $(i,\lambda)$ take values $i=0,\ldots, I$ 
and $\lambda = I + 1, \ldots, h^{1,2}$ for some 
integer $I$, $0\leq I \leq h^{1,2}_+$.   
The only nontrivial intersection numbers are 
\be\label{eq:intno}
\int_Y \alpha_i \wedge \beta^j = \delta_i^j \qquad
\int_Y \alpha_\rho \wedge \beta^\lambda = \delta_\rho^\lambda.
\ee
In terms of this basis of cycles, the holomorphic three-form 
$\omega_Y$ has the following expansion 
\be\label{eq:perexpA} 
\Omega_Y = Z^i\alpha_i+iZ^\lambda \alpha_\lambda - \CF_\lambda \beta^\lambda 
-i\CF_i \beta^i. 
\ee
The orientifold projection imposes the following constraints on the 
periods 
\be\label{eq:orper} 
\im(CZ^i)=\re(C\CF_i)=0\qquad 
\re(CZ^\lambda) = \im(C\CF_\lambda) =0 
\ee
where $C$ is the compensator field introduced in 
\cite{Grimm:2004ua}. For completeness recall that 
\[
C = e^{-\Phi-i\theta} e^{K^{cs}/2}
\]
where $\Phi$ is the four-dimensional dilaton, $e^{i\theta}$ is a phase 
defined by 
\[ 
\eta^* \Omega_Y = e^{2i\theta} {\overline \Omega}_Y
\]
and $K^{cs}$ is the \kah\ potential of the underlying $N=2$ theory 
evaluated on the invariant subspace under the orientifold involution. 

In the spirit of homological mirror symmetry, we can choose the 
symplectic basis so that $Z^0$ 
is the $N=2$ central charge of a point-like brane on $X$ and 
$(Z^i,Z^\lambda)$, $i>0$, 
are central charges of branes wrapping holomorphic 
curves in $X$. 
$\CF^0$ can be similarly taken to be the central charge of a 
D6-brane wrapping $X$ and $(\CF_i,\CF_\lambda)$ can be 
identified with central charges of branes wrapping holomorphic 
divisors in $X$. 
Note that there must be a correlation between the transformation 
properties of holomorphic branes under the holomorphic involution
$\sigma :X \to X$ and the reality properties  
\eqref{eq:orper} of the corresponding periods.
Invariant brane configurations should correspond to real periods 
while anti-invariant brane
configurations should correspond to purely imaginary periods. 
In particular, since the IIB orientifold preserves point-like 
branes on $X$, the fundamental period at the IIA large complex 
structure limit must be real with respect the anti-holomorphic 
involution $\eta:Y\to Y$. This implies that the phase $e^{i\theta}$ 
must equal $1$ in such models. Then equations 
\eqref{eq:orper} reduce to 
\be\label{eq:orperB}
\im(Z^i)=\re(\CF_i)=0\qquad 
\re(Z^\lambda) = \im(\CF_\lambda) =0. 
\ee

The holomorphic coordinates on the $N=1$ complex structure moduli space 
are defined by the periods 
\be\label{eq:holcoord} 
\begin{aligned} 
N^i & = {1\over 2} \int_Y \Omega_{Y}^c \wedge \beta^i \cr
T_\lambda & = i\int_Y \Omega_{Y}^c \wedge \alpha_\lambda\cr
\end{aligned}
\ee 
where $\Omega_{Y}^c$ is a linear combination of the RR three-form 
field $C^{(3)}$ and the (real part of) holomorphic three-form $C\Omega_Y$
\[
\Omega_{Y}^c = C^{(3)} + 2i\re(C\Omega_Y).
\]
Note that the periods \eqref{eq:holcoord} actually parameterize an 
$h^{1,2}(X)+1$ moduli space, which includes the expectation value 
of the dilaton field. 

We can make a more specific choice of symplectic 
basis if we impose additional constraints on the model. 
More specifically we will add the following 
two conditions to the list below equation \eqref{eq:orientifA}
\begin{itemize} 
\item[$(iii)$] The natural push-forward maps 
\be\label{eq:push} 
H_2(S) \to H_2(X)\qquad H_2(S') \to H_2(X)  
\ee
have rank one. 
\item[$(iv)$] The anti-invariant subspace $H^{1,1}_-(X)$ is one 
dimensional and spanned by the difference $S-S'$ between the divisor 
classes of the conjugate del Pezzo surfaces $S,S'$. 
\end{itemize} 

\noindent 
Note that condition $(iii)$ can be reformulated as 
\begin{itemize} 
\item[$(iii')$] The natural restriction maps 
\be\label{eq:restr} 
H^2(X)\to H^2(S)\qquad H^2(X)\to H^2(S') 
\ee
have rank one. 
\end{itemize}

\noindent
Since the second homology of a del Pezzo surface is generated 
by effective curves, condition $(iii)$ implies that we can 
pick two effective curve classes $\Sigma,\Sigma'$ on $X$
which generate the images of the maps \eqref{eq:push}. 
Moreover, these curve classes can be chosen so that 
\be\label{eq:intnoA} 
\begin{aligned}
&\Sigma \cdot S = -1& \qquad & \Sigma\cdot S' =0&  \cr
& \Sigma'\cdot S =0&  \qquad & \Sigma'\cdot S =-1 \cr
\end{aligned}
\ee 
Note that we have the following triple intersection numbers in $X$
\be\label{eq:tripleA}
S^3 = (S')^3 = 9-n
\ee 
for $S,S'\simeq dP_n$. Recall that according to condition 
$(i)$ below \eqref{eq:orientifA} $n$ takes values $6,7,8$
in our models. 
Therefore the self-intersections 
$S^2$ and ${(S')}^2$ must be nontrivial curve classes 
on $X$ which have nonzero intersection numbers with 
$S,S'$ respectively. 
Then we can take $\Sigma, \Sigma'$ so that 
\be\label{eq:intnoC} 
S^2 = (n-9) \Sigma \qquad 
{(S')}^2 = (n-9) \Sigma'. 
\ee

Using condition $(iv)$, we can also choose a system of generators 
$\{J_A\}$ of the \kah\ cone of $X$ so that $J_1=S-S'$ generates 
$H^{1,1}_-(X)$ and $J_A$, $A=2,\ldots,h^{1,1}(X)$  generate 
$H^{1,1}_+(X)$. Moreover, we can make this choice so that 
\be\label{eq:intnoB} 
\begin{aligned} 
& (\Sigma- \Sigma') \cdot J_1 = 1,\qquad (\Sigma -\Sigma') \cdot J_A =0,
\qquad A=2,\ldots, h^{1,1}(X)\cr
& (\Sigma + \Sigma') \cdot J_2 =1, \qquad (\Sigma+\Sigma') \cdot J_A = 0,
\qquad A=1,\ldots, h^{1,1}(X),\ A\neq 2.\cr
\end{aligned}
\ee
This follows from \eqref{eq:intnoA} observing that the 
restriction of any divisor class 
$D$ on $X$ to $S$ and $S'$ respectively must be a multiple of the canonical 
classes $K_S$, $K_{S'}$. 

Another consequence of condition $(iv)$ is that the linear space 
spanned by purely imaginary  
periods of the form $\CF_i$ is two dimensional, i.e. $I=1$ and the index 
$i$ takes only two values $i=0,1$. We can choose the generators 
$\CF_0,\CF_1$ to be the period corresponding to a D6-brane wrapping 
$X$ and the period corresponding to the anti-invariant divisor class 
$J^1=S-S'$. We can also choose the remaining basis elements 
so that the real periods $\CF_\lambda$, $\lambda=1,\ldots,h^{1,2}(Y)$ 
correspond to the divisor classes $J^A$, $A=2,\ldots,h^{1,1}(X)$. 
In particular, $\CF_2$ corresponds to the \kah\ cone generator $J^2$
singled out above equation \eqref{eq:intnoB}. 

Then, according to the orientifold constraints \eqref{eq:orperB},
the dual periods $Z^i$, $i=0,1$, must be real and the remaining 
periods $Z^{\lambda}$ must be imaginary. We take $Z^0$ to be  
the period associated to a point-like brane on $X$.
Taking into account the intersection numbers \eqref{eq:intnoB}, 
we can choose the basis of cycles so that $Z^1,Z^2$ are 
associated to the curve classes $\Sigma-\Sigma'$, 
$\Sigma+\Sigma'$. 
This is consistent with the fact that the O3 orientifold projection 
maps a D5-brane supported on $\Sigma$ to an anti-D5-brane supported 
on $\Sigma'$. 

As discussed in the previous section, the type IIB fractional 
associated to a collapsing del Pezzo surface in $X$ are related by mirror 
symmetry to a collection of D6-branes wrapping special lagrangian 
cycles in $Y$. The IIB fractional branes associated to the 
del Pezzo surface $S$ carry D3, D5 and D7 charges. We have 
to work out the corresponding D6-brane charges on the mirror 
threefold $Y$ in terms of the basis of cycles 
$(\alpha_i,\alpha_\lambda; \beta^i,\beta^\lambda)$.
 
Note that in the compact model we have two fractional 
brane configurations mapped to each other by the orientifold 
projection. Suppose we have a fractional brane $E_a$ on $S$
with topological charges 
\be\label{eq:fractchargesA}
r_a S + n_a \Sigma + m_a \omega,
\ee 
where $\omega$ is the class of a point on $X$.  
The orientifold projection will map it to a fractional 
brane $E'_a$ supported on $S'$ with charge vector 
\be\label{eq:fractchargesB} 
r'_a S'+ n'_a \Sigma' + m'_a \omega 
\ee
where 
\[ 
r'_a = r_a \qquad n'_a = -n_a \qquad m'_a= m_a.
\]
Since the del Pezzo surfaces $S,S'$ are conjugated under the holomorphic
involution, $S+S'$ is an invariant divisor class on $X$. Therefore 
it corresponds by mirror symmetry to an invariant 
period $\CF_+$ which can be written as linear combination 
\be\label{eq:fractchargesC} 
\CF_+ = \sum_{\lambda=2}^{h^{1,2}(Y)} s^\lambda \CF_\lambda 
\ee
with integral coefficients. Since the anti-invariant combination
$S-S'$ is related by mirror symmetry to $\CF^1$, and $Z^1,Z^2$ are related 
to the curve classes $\Sigma\pm \Sigma'$, we find 
\be\label{eq:chcharges} 
\begin{aligned} 
Z(E_a) & = {1\over 2}r_a (\CF_+ + \CF^1) + {1\over 2}
n_a (Z^2+Z^1) + m_a Z^0 \cr
Z(E'_a) & = {1\over 2}r_a (\CF_+-\CF^1) - {1\over 2}n_a (Z^2-Z^1) 
+ m_a Z^0\cr
\end{aligned} 
\ee
Note that the linear combinations \eqref{eq:chcharges} 
have half integral coefficients because the basis of cycles 
$(\alpha_i,\alpha_\lambda; \beta^i,\beta^\lambda)$ we have
constructed above generates $H^3(Y)$ over $\IQ$, but not over 
$\IZ$. In fact in the models considered here it is impossible 
to find a basis over $\IZ$ consisting of invariant and anti-invariant 
cycles. 

Equations \eqref{eq:chcharges} determine the charges of the D6-brane 
configurations on $Y$ related by mirror symmetry to the fractional 
branes $(E_a,E'_a)$. We have 
\be\label{eq:mirrorfract} 
\begin{aligned}
& M_a=\half r_a(s^\lambda\alpha_\lambda+\alpha_1) + 
\half n_a (\beta^2+\beta^1) 
+ m_a \beta^0\cr
& M'_a=\half r_a(s^\lambda\alpha_\lambda -\alpha_1) -\half n_a 
(\beta^2-\beta^1)  +m_a \beta^0\cr
\end{aligned} 
\ee

We will be interested in fractional brane configurations 
of the form $\oplus_{a=1}^s E_a^{\oplus N_a}$ which give rise 
to supersymmetry breaking quivers. 
Note that the total D7-charge of such a configuration vanishes 
\be\label{eq:dseven} 
\sum_{a} N_a r_a = 0. 
\ee
The total D5-brane charge is represented by a curve class 
$\Gamma$ on $S$ which is orthogonal to the canonical class 
$K_S$ i.e. 
\be\label{eq:orth} 
(\Gamma \cdot K_S)_S =0.
\ee
Condition $(iii)$ above implies then that $\Gamma$ must be contained 
in the kernel of the pushforward map $H_2(S)\to H_2(X)$, hence 
$\Gamma$ is homologically trivial in $X$, and  
the total D5-brane charge vanishes as well. 
This is also valid for the conjugate fractional 
brane configuration supported on $S'$. 
The total D3-brane may be nonzero, depending on the choice 
of fractional brane configuration. More precisely, the total 
D3-brane charge for the supersymmetry breaking quiver 
represented in fig. 1 is an integer $N$, which  
is constrained by tadpole cancellation and supersymmetry. 
The total D3-brane charge, that is the charge $N$ of the fractional 
branes plus the charge $N_{D3}$ of additional D3-branes on $X$ must equal 
the absolute value $|N_{O3}|$ of the total charge of the O3-planes.  
Moreover, we do not allow anti-D3-branes on $X$ in order to preserve 
tree level supersymmetry. 
In addition we may also have D7-branes wrapping the
divisors in $X$ if the holomorphic involution 
$\sigma$ has codimension one fixed loci. 

Note that this situation is different from the D-brane 
landscape constructed in \cite{DGS:land}. The model 
constructed in \cite{DGS:land} included a D5 anti-D5 
pair wrapping conjugate curves in a Calabi-Yau threefold. 
This eventually led to tree level supersymmetry breaking 
while in the present case supersymmetry breaking occurs 
as a result of strong infrared dynamics at a much lower scale.  

In IIA compactifications, complex structure moduli can be stabilized 
by turning on NS-NS flux. The most general 
flux compatible with the orientifold projection is given by
\cite{Grimm:2004ua} 
\be\label{eq:nsnsfluxA}
H= q^\lambda\alpha_\lambda - p_k \beta^k 
\ee 
where $q^\lambda, p_k$ are integers. 
According to \cite{Grimm:2004ua}, the flux superpotential is 
\be\label{eq:suppotB} 
W_H = -2 N^k p_k -i T_\lambda q^\lambda. 
\ee
Such a superpotential would generically fix all complex structure 
moduli and the dilaton. 

However in the presence of branes the flux \eqref{eq:nsnsfluxA}
is subject to the Freed-Witten anomaly cancellation  condition
\eqref{eq:FWanom}. In our situation, the three-cycles supporting 
fractional brane charges can be read off from equations 
\eqref{eq:mirrorfract}. 
We obtain the following constraints
\be\label{eq:FWanomB} 
\int_X H \wedge \gamma =0, \qquad \gamma=\beta^0,\beta^1,\beta^2,
\alpha_1,s^\lambda\alpha_\lambda.
\ee 
Using the intersection numbers \eqref{eq:intno}, we find that 
the conditions \eqref{eq:FWanomB} are equivalent to 
\be\label{eq:FWanomC} 
p_1=q^2=0.
\ee
Therefore the superpotential \eqref{eq:suppotB} does not depend on the
moduli fields $N^1,T_2$, 
but it has nontrivial dependence on all the other complex 
structure moduli and the dilaton. 

In principle, we can have additional constraints corresponding 
to background D3 and D7-branes on $X$ supported away from the 
del Pezzo surfaces $S,S'$. Any D3-brane on $X$ is mapped by mirror 
symmetry to a D6-brane with charge $\beta^0
\in H^3(Y)$. Therefore the corresponding 
Freed-Witten constraint has already been taken into account in 
\eqref{eq:FWanomC}. 
A D7-brane wrapping an invariant divisor $D$ is mapped by 
mirror symmetry to 
D6-brane wrapping a special lagrangian cycle $M$ in $Y$
so that $b_1(M) = h^{0,2}(D)$. 
Since $D$ is invariant under the 
holomorphic involution, the class of the mirror cycle 
$M$ will be a linear combination of basis elements of the form 
\be\label{eq:mixedA} 
M = \sum_{\lambda=2}^{h^{1,2}(Y)} d^\lambda \alpha_\lambda. 
\ee
Taking into account the intersection numbers \eqref{eq:intno} 
and equation \eqref{eq:nsnsfluxA}, it follows 
that the presence of D6-branes wrapping $M$ does not yield additional 
Freed-Witten constraints on the background flux $H$. 

One may worry however about open string moduli corresponding 
to normal deformations of the special lagrangian cycle 
$M$ in $Y$. General supersymmetry considerations suggest that  
these moduli should be lifted in the presence of background flux. 
In the present case, this effect has been confirmed by a detailed 
mathematical analysis in \cite{Koerber:2006hh}. Similar results 
in IIB theory and M-theory include 
\cite{Saulina:2005ve,Kallosh:2005gs,Bergshoeff:2005yp}. 
In fact, as long as the cycle $M$ is not very close to the 
cycles $M_a$ supporting the fractional D6-branes, there will 
be no low energy couplings between these brane configurations. 
Therefore dynamical supersymmetry breaking on the fractional 
D6-branes will not be affected by the presence of open 
string moduli on the second stack of D6-branes. 

Summarizing the above discussion, 
the total superpotential for the closed string moduli reads  
\be\label{eq:suppotC}
W = W_F + W_H 
\ee
where $W_F$ is given by \eqref{eq:superpotA} and 
\be\label{eq:suppotD} 
W_H = -2N^0p_0 - i \sum_{\lambda =3}^{h^{1,2}} T_\lambda q^\lambda.
\ee
This yields the following supergravity F-flatness equations 
\be\label{eq:FflatA}
\begin{aligned} 
D_{t^\alpha} W & =0 \qquad &  
& D_{N^0}W & =0 \qquad & & \partial_{N^1} \CK =0\quad & \cr
 {} & \ \ \qquad & & \partial_{T_2} \CK & =0 \qquad& &  
D_{T_\lambda} W =0 \quad& &  \lambda\geq 3.\cr
\end{aligned}
\ee
where $\alpha = 1,\ldots, h^{1,1}(Y)$, $t^\alpha$ are holomorphic 
coordinates on the \kah\ moduli space, and $\CK$ denotes the 
\kah\ potential for the complex structure moduli space.

The number of equations in \eqref{eq:FflatA} equals the number 
of variables, hence one may be tempted to naively conclude that 
this system will generically have isolated solutions in the 
moduli space. However this conclusion would not be reliable without 
a more detailed analysis. A typical problem in such situations 
is that a nongeneric superpotential can leave some flat directions 
unlifted in the moduli space. This is well known for 
IIB flux compactifications \cite{KKLT}, which exhibit a no-scale 
\kah\ potential on the $\CN=1$ \kah\ moduli space,
in the absence of instanton corrections to the superpotential.

In our case the F-flatness equations 
\be\label{eq:FflatB} 
\partial_{N^1} \CK =0\qquad
\partial_{T_2} \CK=0
\ee
may cause in principle a similar problem due to the absence of 
superpotential terms. In order to address this question, we will rewrite 
these two equations using some identities proved in \cite{Grimm:2004ua}. 
Equation (B.9) in \cite{Grimm:2004ua} yields 
\be\label{eq:FflatC} 
\partial_{T^2} \CK = -2 e^{2\Phi}\im(CZ^2),
\ee
where $\Phi$ is the four dimensional dilaton. 
Moreover, combining equations (B.9) and (C.12) in \cite{Grimm:2004ua}, 
we obtain 
\be\label{eq:FflatD} 
\partial_{N^1} \CK = 8 e^{2\Phi} \im(C\CF_1).
\ee
Since in our case $C$ is real, the flatness equations 
\eqref{eq:FflatB} reduce to 
\be\label{eq:FflatE} 
\im(Z^2)=0 \qquad \im(\CF_1)=0. 
\ee
These are the defining equations of the intersection locus between 
the totally real subspace 
of the $N=2$ moduli space fixed by the anti-holomorphic involution 
and the subspace 
\be\label{eq:FflatF} 
Z^2=0\qquad \CF_1=0. 
\ee
We claim that the locus cut out by equations \eqref{eq:FflatF} 
and the quiver locus intersect at least along a codimension two subspace of 
the complex structure moduli space. 
Note that this is a very important feature of the 
F-flatness equations \eqref{eq:FflatA}, since it shows that low 
energy dynamical supersymmetry breaking is generic in 
these models. If the subspace cut by equations \eqref{eq:FflatF} 
intersected the quiver locus along a higher codimension locus, 
dynamical supersymmetry breaking would be non-generic. In that case, 
one would have to fine tune the flux parameters in order to 
obtain flux vacua endowed with supersymmetry 
breaking quivers. Moreover, such a
fine tuning may not be ultimately possible since the 
flux parameters are discrete. 

The claim made below equation \eqref{eq:FflatF} follows from 
certain universal properties of contractible del Pezzo surfaces 
in $\CN=2$ compactifications. 
Recall that we have restricted 
ourselves to del Pezzo surfaces $S,S'\simeq dP_n$, $n=6,7,8$ 
embedded in the Calabi-Yau threefold $X$ so that the 
restriction maps $H^2(X) \to H^2(S)$,  $H^2(X)\to H^2(S')$ 
have rank one. 
Then the properties of del Pezzo fractional branes -- in particular the 
relations between their central charges -- are captured by a
universal one parameter local model derived in 
\cite{Klemm:1996hh,Mayr:1996sh,Lerche:1996ni}. The complete solutions
of these models for $n=6,7,8$,
including analytic continuation to the quiver loci, 
have been worked out in \cite{Mohri:2000kf}. 
Based on the results obtained there, and the 
universal character of local models, we conjecture the following 
properties of the compact models. 
\begin{itemize}
\item[$(a)$] The quiver
locus contains the codimension two subspace of the moduli space where 
the quantum volumes of the curves $\Sigma, \Sigma'$ vanish\footnote{This 
condition may seem at odds with our previous claim that the SCFT 
is nonsingular along the quiver locus. In fact there is no contradiction 
here since although the central charges of $\Sigma,\Sigma'$ vanish, 
there are no stable massless states along this subspace of the 
moduli space. The stable supersymmetric D-branes wrapping 
$\Sigma,\Sigma'$ undergo decay into fractional branes 
along a marginal stability wall which separates the 
LCS point from the quiver locus \cite{Douglas:2000qw}.}.
\item[$(b)$] The central charges of all fractional branes supported 
on a collapsing del Pezzo surface are equal to the central charge 
of a pointlike brane along the locus specified at $(a)$. 
\end{itemize}
We will confirm properties $(a)$, $(b)$
by direct computations for a concrete compact model in the next 
section. 

Note that the subspace where the quantum volumes of $\Sigma,\Sigma'$ 
vanish can be equivalently characterized by the equations 
\be\label{eq:quiverlocA} 
Z^1=0\qquad Z^2=0
\ee
in the symplectic basis of periods adapted to the orientifold projection. 
Moreover property $(b)$ above implies that 
the central charge of a D-brane wrapping $S$ and the central charge 
of a D-brane wrapping $S'$ will be both equal to the central charge 
of a pointlike brane on $X$ along the subspace \eqref{eq:quiverlocA}. 
By construction, the period $\CF_1$ can be written as the difference 
\be\label{eq:difference} 
\CF_1 = Z_S - Z_{S'}
\ee
between the $N=2$ central charges of a D-brane wrapping $S$ and 
a D-brane wrapping $S'$. Therefore $\CF_1$ vanishes identically along the 
subspace \eqref{eq:quiverlocA}. 
This concludes our argument.

\section{Concrete Examples}

In this section we construct examples of Calabi-Yau threefolds $X$ satisfying 
conditions $(i)-(iv)$ imposed in the previous section  and check 
the properties $(a),(b)$ of the quiver locus in a specific model. 

\subsection{An Elliptic Fibration} 

Consider a smooth Weierstrass model ${\widetilde X}$ 
over the del Pezzo surface $B=dP_2$, which is the 
two-point blow-up of the projective plane $\IP^2$. 
Let $p$ and $p'$ denote the centers of the 
blow-ups on $\IP^2$, and let $e$ and $e'$ denote the exceptional curves. 
If the fibration is generic, the restriction 
of the elliptic fibration to any $(-1)$ curve $C$ on $B$ is 
isomorphic to a rational elliptic surface with $12$ $I_1$ fibers, 
usually denoted by $dP_9$. Therefore, taking $C$ to be each of the 
two exceptional curves, we obtain two rational elliptic
surfaces denoted by $D$ and $D'$. 
The exceptional curves $e,e'$ can be naturally identified
with sections of the rational elliptic surfaces $D,D'$. 
We can also naturally regard them as
$(-1,-1)$ curves on ${\widetilde X}$, 
by embedding $B$ in ${\widetilde X}$ via the section of the Weierstrass model.

Now, we can perform a flop on the $(-1,-1)$ curves 
$e,e'$ in ${\widetilde X}$, obtaining an elliptic fibration $\pi:X
\to\IP^2$ with two complex dimensional components in the fiber. 
More precisely, the fibers over the points $p,p'$ have 
two components: a rational $(-1,-1)$ curve obtained by 
flopping one of the curves $e,e'$, denoted by $C$ and $C'$ respectively, 
and a $dP_8$ del Pezzo surface. 
We will denote the $dP_8$ components by $S$ and $S'$. 
It is then possible to contract 
the del Pezzo surfaces in the fiber, obtaining a singular elliptic 
fibration ${\widehat X}$ over $\IP^2$, which can be eventually 
smoothed out by complex structure deformations \cite{Morrison:1996pp}. 

The Calabi-Yau $X$ has Hodge numbers $(h^{1,1}( X),h^{2,1}(X))=(4,214)$. 
The Mori cone of $X$ is given by:
\be\label{eq:Mcone}
\begin{array}{ccccccccc}
& W_1 & W_2 & W_3 & W_4 & W_5 & Z & X & Y \cr
\Sigma: & 1 & 0 & 1 & -1 & 0 & 0 & 2 & 3 \cr
\Sigma':& 0 & 1 & 1 & 0 & -1 & 0 & 2 & 3\cr
C: & -1 & 0 & -1 & 1 & 0 & 1 & 0 & 0 \cr
C':& 0 & -1 & -1 & 0 & 1 & 1 & 0 & 0 \cr
h: & 1 & 1 & 1 & 0 & 0 & -3 & 0 & 0, \cr
\end{array}
\ee
where $W_4$ and $W_5$ are the rays in the toric fan of the ambient 
space which correspond to the del Pezzo $dP_8$ surfaces $S$ and $S'$ 
respectively. $h$ denotes the hyperplane class of $\IP^2$ and 
$\Sigma,\Sigma'$ are the generators of the images of the 
restriction maps \eqref{eq:restr}. 
The disallowed locus is given by
\be\label{eq:forbidden} 
\{W_3=W_4=0\} \cup \{W_2=W_5=0\}\cup \{W_4=W_5=0\} \cup\{W_1=W_2=W_3=0\}.
\ee
Note that the Mori cone is not simplicial and we have to chose a simplicial 
subcone in order to write down the mirror Picard-Fuchs equations.
This choice is not essential, since it corresponds to a choice of 
large complex limit coordinates on the moduli space.
In the following we will work with the simplicial subcone generated 
by $(\Sigma,\Sigma',C,h)$. 

The IIB orientifold is defined by the following involution:
\be\label{eq:holinvC}
\sigma :(W_1,W_2,W_3,W_4,W_5,Z,X,Y)\to (W_1,W_3,W_2,W_5,W_4,Z,X,Y),
\ee
which does preserve the large complex structure limit. Taking into account the toric data \eqref{eq:Mcone}, 
the fixed point set equations are given by
\be\label{eq:fps}
\begin{aligned}
&W_1=\lb_1\lb_2\lb_3^{-1}\lb_4W_1, & 
& W_2=\lb_1\lb_3^{-1}\lb_4W_3,& & W_3=\lb_2\lb_4W_2, &\\
&W_4=\lb_1^{-1}\lb_3W_5,& & W_5=\lb_2^{-1}W_4,& & Z=\lb_3\lb_4^{-3}Z,&\\
&X=\lb_1^2\lb_2^2X,& & Y=\lb_1^3\lb_2^3Y, & & &\\
\end{aligned}
\ee
where $\lb_i\in \IC^*$, $i=1,\ldots,4$. 
The second and third equations in \refeq{fps} yield
\be\label{eq:fpsi}
W_2W_3=\lb_1\lb_2\lb_3^{-1}\lb_4^2W_2W_3
\ee
while the fourth and fifth equation in \refeq{fps} give
\be\label{eq:fpsii}
W_4W_5=\lb_1^{-1}\lb_2^{-1}\lb_3W_4W_5.
\ee
But $W_4$ and $W_5$ are not allowed to vanish simultaneously, 
therefore we must have
\be\label{eq:fpsiii}
\lb_1^{-1}\lb_2^{-1}\lb_3=1.
\ee
Then, the fourth and fifth equations in \refeq{fps} reduce to
\be\label{eq:fpsiv}
W_4=\lb_2W_5.
\ee
Substituting \refeq{fpsiii} in \refeq{fpsi} and taking into account the 
first equation in \refeq{fps}, we obtain
\be\label{fpsv}
W_1=\lb_4W_1 \qquad W_2W_3=\lb_4^2W_2W_3.
\ee
Since $W_1,~W_2,~W_3$ are not allowed to vanish simultaneously, 
taking also into account \refeq{fpsiv}, we obtain two cases
\be\label{eq:fixedKi} 
\begin{aligned}  
\lambda_4=1 \ & \Rightarrow \ W_3=\lambda_2 W_2,\  W_4 = \lambda_2 W_5 \cr
\lambda_4=-1 \ & \Rightarrow \ W_3 = -\lambda_2 W_2, \ W_1=0, \
W_4=\lambda_2 W_5. \cr
\end{aligned} 
\ee 
We can eliminate $\lb_2$ from the above equations, obtaining
the following cases:
\be\label{fpsvi}
\begin{aligned}
\lambda_4=1 \ & \Rightarrow \ W_2W_4=W_3W_5\cr
\lambda_4=-1 \ & \Rightarrow \ W_2W_4=-W_3W_5, \ W_1=0.\cr
\end{aligned}
\ee
If $\lb_4=1$, the remaining equations in \refeq{fps} yield
\be\label{fpsvii}
X=\lb_3^2X,\quad Y=\lb_3^3Y,\quad Z=\lb_3Z.
\ee
and therefore the whole divisor $D$ given by the equation 
$W_2W_4-W_3W_5=0$ in the Calabi-Yau threefold $X$ is fixed by the 
holomorphic involution $\sigma $. 
Note that $D$ does not intersect the del Pezzo surfaces $S$ and $S'$.

If $\lb_4=-1$, we obtain
\be\label{fpsviii}
X=\lb_3^2X,\quad Y=\lb_3^3Y,\quad Z=-\lb_3Z,
\ee
which can only be satisfied if $Z=0$ or $Y=0$. 
Taking into account that in this case we also have 
$W_1=0$ and $W_2W_4+W_3W_5=0$, 
we conclude that this component of the fixed locus 
consists of a finite set of points. 
Note that these points are away from $D$ as 
well as the del Pezzo surfaces $S$ and $S'$.

Next we will show that the generic hypersurface $X$ preserved by the 
involution \eqref{eq:holinvC} is smooth. 
It suffices to show that ${\widetilde X}$ 
is nonsingular, since $X$, ${\widetilde X}$ are related by
a flop. ${\widetilde X}$ is a hypersurface in a toric fourfold $P$
which is a $\IP^2_{[1,2,3]}$-bundle over $dP_2$. 
This is closely related to the $\IP^2_{[1,2,3]}$-bundle $P'$ over $\IP^2$, 
which is defined by the following toric data  
\be\label{eq:tordata}
\begin{array}{lllllll}
& W_1 & W_2 & W_3 &  Z &  X &  Y\\
& 1  & 1  & 1 & -3 & 0 & 0 \\ 
& 0 & 0 & 0 & 1 & 2 & 3.\\ 
\end{array}
\ee 
The transition from
$P'$ to $P$ can be accomplished by adding two extra homogeneous 
variables $W_4, W_5$ and modifying the torus action accordingly. 
Equivalently, we can work over the base $\IP^2$ (i.e. on
the fourfold $P'$) but require that all sections vanish to appropriate
order at the two blown-up points 
\[p_1=[0,1,0],\quad p_2=[0,0,1].\]
Thus, for
example the anticanonical bundle has 10 sections on $\IP^2$, namely the
cubic
monomials in the $W_i$, $i=1,2,3$, but only 8 
of them survive on $dP_2$ - eliminate
$W_2^3$ and $W_3^3$. The most general Calabi-Yau hypersurface in $P'$ is
given by a combination of monomials of the form:
\[
Y^2,\ a_3(W)XYZ,\ a_9(W)YZ^3,\ X^3,\ a_6X^2Z^2,\ a_{12}(W)XZ^4,\ a_{18}(W)Z^6.
\]
Here $a_{3i}$ is a monomial of degree $3i$, $i=1,\ldots,6$ 
in $W_1,W_2,W_3$. The analogous
statement for $P$ is obtained by taking $a_{3i}$ to be a monomial of
degree $3i$ which vanishes to order $i$ at $p_1$ and $p_2$, $i=1,\ldots,6$. 
The "Weierstrass" sublinear system
is generated only by the polynomials:
\be\label{eq:weierstrass} Y^2-X^3,\ a_{12}(W)XZ^4,\ a_{18}(W)z^6.\ee
These hypersurfaces have a zero section: $Z=0,\ X=1,\ Y=1$. As explained 
above, we
want to consider only the sublinear system consisting of Weierstrass
hypersurfaces which are also invariant under the involution \eqref{eq:holinvC}.
Since $Z$, $X$ and $Y$ are invariant, this simply means that the
coefficients $a_{12}$, $a_{18}$ are symmetric in $W_2$ and $W_3$.
Bertini's theorem allows us to conclude that a generic hypersurface in
a linear system is smooth without actually exhibiting such a smooth
hypersurface. It asserts that a generic hypersurface can be singular
only at points of the base locus.
In order to apply Bertini in our case, we need the base locus of 
the monomials \eqref{eq:weierstrass}. This is
clearly just the zero section: the three symmetric polynomials
\[(W_1^{18})Z^6,\ (W_2^{18}+W_3^{18})Z^6,\ (W_2W_3Z)^6\]
already cut out the divisor $Z=0$, which together with $Y^2-X^3$ gives the
zero section. So a generic $\sigma$-invariant Weierstrass hypersurface is
smooth away from its zero section. But since sections of an elliptic
fibration cannot pass through singular points, it is smooth at points
of the zero section as well. 

Now, we 
claim that the model $(X,\sigma)$ constructed above satisfies conditions 
$(i),(ii)$ below \eqref{eq:orientifA} as well as conditions 
$(iii), (iv)$ below \eqref{eq:holcoord}. Condition $(i)$ follows easily 
from the definition of the holomorphic involution \eqref{eq:holinvC}, 
which interchanges the defining equations $W_4=0$ and $W_5=0$ of the 
two del Pezzo surfaces respectively. Condition $(ii)$ follows from the 
fact that the holomorphic involution $\sigma$ is a symmetry of the toric 
polytope of $X$. The toric polytope of $X$ is identified with the dual 
toric polytope of $Y$ under the monomial-divisor map \cite{Aspinwall:1993nu,
Aspinwall:1993rj}. It is clear that the large radius limit in the 
complexified \kah\ moduli space can be reached by increasing the sizes 
of the toric divisors of $Y$ in an invariant fashion with respect to the 
given involution. Then the monomial-divisor map implies that the 
large complex structure of $X$ can be reached in a similar fashion 
in the complex structure moduli space. 
Conditions $(iii)$ and $(iv)$ follow from a 
simple moduli count. We know that $h^{1,1}(X)=4$ and we have at least two 
independent divisor classes -- a vertical divisor class $\pi^*(h)$ and 
the section class $B$ -- on $X$. The two del Pezzo surfaces $S,S'$ can be 
contracted independently on $X$ 
(this can be achieved by toric contractions) therefore they provide two 
additional independent generators of the Picard group. 
Taking into account the disallowed locus \eqref{eq:forbidden}, 
it is straightforward to check that the only generator of the 
Picard group which restricts nontrivially to $S$ is $S$ itself. 
The same statement holds for the conjugate del Pezzo surface 
$S'$. By construction, the divisor classes $\pi^*(h), B, S+S'$ are 
invariant under the holomorphic involution $\sigma$ while 
$S-S'$ is anti-invariant. Therefore conditions 
$(iii)$, $(iv)$ are indeed satisfied. 

Checking properties $(a)$, $(b)$ of the quiver locus listed above 
\eqref{eq:quiverlocA} requires a more involved computation. 
In particular we have to solve the Picard-Fuchs equations for 
the mirror threefold $Y$ and identify the region of the 
moduli space where the quantum volumes of the curves $\Sigma, 
\Sigma'$ vanish. 
Using the simplicial subcone of the Mori cone generated by 
$(\Sigma,\Sigma',C,h)$, we find that the K\"ahler form of $X$ is given by
$$
J=t_1(3W_1+2W_4+3W_5+Z)-t_2W_5+t_3(3W_1+3W_4+3W_5+Z)+t_4(W_1+W_4+W_5).
$$ 
The large radius prepotential on the K\"ahler moduli space of $X$ reads
\be\label{eq:Prepo}
\begin{aligned}
{\cal F}_{\cal K}^X=&\frac{4}{3}t_1^3-\frac{1}{6}t_2^3+\frac{3}{2}t_3^3
+\frac{9}{2}t_1^2t_3+\frac{9}{2}t_1t_3^2+\frac{3}{2}t_1^2t_4+
\frac{1}{2}t_1t_4^2
+\frac{3}{2}t_3^2t_4+\frac{1}{2}t_3t_4^2+3t_1t_3t_4-\frac{23}{6}t_1+
\frac{5}{12}t_2\cr
&-\frac{3}{2}t_4-\frac{17}{4}t_3+\frac{105i}{4\pi^3}\zeta(3)+{\cal O}(q_i),
\end{aligned}
\ee
where $q_i=e^{t_i/2\pi i}$, $i=1,\ldots,4$.
In the vicinity of the large complex structure point, the 
fundamental period is given by:
\be
\begin{aligned}
X^0=
\sum_{n_1,n_2,n_3,n_4\geq 0}&\frac{z_1^{n_1}z_2^{n_2}z_3^{n_3}z_4^{n_4}}
{\Gamma(1+n_1-n_3+n_4)\Gamma(1+n_2+n_4)\Gamma(1+n_1+n_2-n_3+n_4)
\Gamma(1-n_1+n_3)}\cr
&\times\frac{\Gamma(1+6n_1)\Gamma(1+6n_2)}{\Gamma(1-n_2)
\Gamma(1+n_3-3n_4)\Gamma(1+2n_1+2n_2)\Gamma(1+3n_1+3n_2)},
\end{aligned}
\ee
where $z_1,\ldots,z_4$ are algebraic coordinates on the moduli space 
centered at the large complex structure limit point. The quantum volumes 
of $\Sigma, \Sigma'$ are given by the logarithmic 
periods 
\be\label{eq:logper}
\begin{aligned}
X^1=&\frac{1}{2\pi i}X^0\ln z_1+\frac{1}{2\pi i}g_1,\cr
X^2=&\frac{1}{2\pi i}X^0\ln z_2+\frac{1}{2\pi i}g_2,
\end{aligned}
\ee
where
\be
\begin{aligned}
g_1=\sum_{n_1,n_2,n_3,n_4}\Big[&h_0(6S_{n_1}-S_{n_1-n_3+n_4}-
S_{n_1+n_2-n_3+n_4}+S_{-n_1+n_3}-2S_{2n_1+2n_2}-3S_{3n_1+3n_2})\cr
&-h_1-h_2+h_3\Big],\cr
g_2=\sum_{n_1,n_2,n_3,n_4}\Big[&h_0(6S_{n_2}-S_{n_2+n_4}-
S_{n_1+n_2-n_3+n_4}+S_{n_2}-2S_{2n_1+2n_2}-3S_{3n_1+3n_2})-h_2+h_4\Big].\cr
\end{aligned}
\ee
In the above expressions we have used the notation 
$S_n=\sum_{k=1}^n\frac{1}{k}$; the expressions of the functions 
$h_i(z_1,\ldots,z_4)$, $i=0,\ldots,4$ are presented in appendix A.
The periods $Z^1,Z^2$ defined in section three are related to $X^1,X^2$ by 
$$
Z^1=X^1-X^2,\quad Z^2=X^1+X^2.
$$

We claim that the periods $X^1,X^2$ vanish simultaneously along the 
subspace $\CQ$ in the complex structure moduli space of $Y$ 
defined by $z_1=z_2=\infty$. This locus lies outside the domain of 
convergence of the large complex structure periods, therefore we 
have to perform analytic continuation. Following the prescription 
of \cite{Aspinwall:1994zu}, the good algebraic coordinates 
in a neighborhood of  ${\cal Q}$ are 
$$
w_1=\frac{1}{z_1},~~w_2=\frac{1}{z_2},~~w_3=z_1z_3,~~w_4=z_4.
$$ 
Next, note that the fundamental period $X^0$ does not actually 
require analytical continuation. The above change of variables yields a 
convergent series 
\be\label{eq:newper}
\begin{aligned}
{\widetilde X}^0=\sum_{p_3\geq p_1\geq 0,p_4\geq 0}&
\frac{w_1^{p_1}w_3^{p_3}w_4^{p_4}}{\Gamma(1-p_1+p_4)^2
\Gamma(1+p_4)\Gamma(1+p_1)
\Gamma(1+p_3-p_4)\Gamma(1-2p_1+2p_3)}\cr
&\times\frac{\Gamma(1-6p_1+6p_3)}{\Gamma(1-3p_1+3p_3)}.
\end{aligned}
\ee
Note also that 
${\widetilde X}^0$ has no zeroes on ${\cal Q}$
since the zero-th order term in the expansion \eqref{eq:newper} 
is 1. 

The logarithmic periods however require analytic continuation. 
Representing the 
periods as Barnes integral and deforming the integration contour, 
we obtain, using the prescription of \cite{Aspinwall:1994zu}:
\be\label{eq:lognewper}
\begin{aligned}
{\widetilde X}^1=&-\frac{1}{2\pi i}\left[
\frac{\Gamma\left(-\frac{1}{6}\right)
\Gamma\left(\frac{1}{6}\right)}{6\sqrt{\pi}
\Gamma\left(\frac{2}{3}\right)\Gamma\left(\frac{5}{6}\right)}w_1+ 
\frac{\Gamma\left(-\frac{7}{6}\right)\Gamma\left(\frac{7}{6}\right)}
{6\sqrt{\pi}
\Gamma\left(-\frac{1}{6}\right)\Gamma\left(\frac{2}{3}\right)}
w_1^2w_3+\cdots\right],\cr
{\widetilde X}^2=&-\frac{1}{2\pi i}\left[\frac{64\Gamma
\left(-\frac{1}{6}\right)\Gamma\left(\frac{1}{6}\right)}{\sqrt{\pi}
\Gamma\left(\frac{8}{3}\right)\Gamma\left(\frac{5}{6}\right)}
w_2w_3+160w_2^3w_3+\cdots\right].
\end{aligned}
\ee
In order to prove property $(b)$ we also have to compute 
the central charges of D4-branes wrapped on the 
del Pezzo surfaces $S$ and $S'$ in a neighborhood of $\CQ$.  
This requires analytic continuation of the quadratic logarithmic 
periods near the large radius limit. 
Using \refeq{Prepo}, we can write a basis for the quadratic periods 
of the form 
\be
\begin{aligned}
{\cal G}_1=&
-\frac{X^0}{4\pi^2}\left[4(\ln z_1)^2+\frac{9}{2}(\ln z_3)^2+
\frac{1}{2}(\ln z_4)^2+9\ln z_1\ln z_3+3\ln z_1\ln z_4+
3\ln z_3\ln z_4\right]\cr
&-\frac{23}{6}X^0+\ldots,\cr
{\cal G}_2=&\frac{X^0}{8\pi^2}(\ln z_2)^2+\frac{5}{12}X^0+\ldots,\cr
{\cal G}_3=&
-\frac{X^0}{4\pi^2}\left[\frac{9}{2}(\ln z_1)^2+\frac{9}{2}(\ln z_3)^2
+\frac{1}{2}(\ln z_4)^2+9\ln z_1\ln z_3+
3\ln z_1\ln z_4+3\ln z_3\ln z_4\right]\cr
&-\frac{17}{4}X^0+\ldots,\cr
{\cal G}_4=&
-\frac{X^0}{4\pi^2}\left[\frac{9}{2}(\ln z_1)^2+
\frac{9}{2}(\ln z_3)^2+3\ln z_1\ln z_3+\ln z_1\ln z_4+\ln z_3\ln z_4\right]
-\frac{3}{2}X^0+\ldots,\cr
\end{aligned}
\ee  
where the dots stand for holomorphic terms.
Near the large radius limit, the central charges 
$Z_S,Z_{S'}$
have an expansion 
\be\label{eq:lcsexpA}
\begin{aligned} 
Z_S & = Z^0 \left[\int_X e^{J}\mathrm{ch}(\CO_S) \sqrt{\mathrm{Td(X)}} 
+ \ \mathrm{instanton \ corrections}\ \right]\cr
Z_{S'} & = Z^0 \left[\int_X e^J\mathrm{ch}(\CO_{S'}) 
\sqrt{\mathrm{Td(X)}} + \ \mathrm{instanton \ corrections}\ \right]\cr
\end{aligned}
\ee
where $J$ is the \kah\ form on $X$ expressed in terms of the flat 
coordinates $t_1,\ldots,t_4$. Note that for Calabi-Yau threefolds 
\[
\sqrt{\mathrm{Td(X)}}= 1+ {1\over 24}~ c_2(X).
\]
For the present model, the second Chern class is given by
$$
c_2(X)=102W_1^2+137W_1W_4+45W_4^2+137W_1W_5+45W_5^2+
92W_4W_5+69W_1Z+46W_4Z+46W_5Z+11Z^2.
$$
Taking into account the triple intersection numbers
\be
\begin{aligned}
& W_1^3=-2& & W_4^3=1& &W_5^3=1& & &\\
&W_1^2W_4=1 & & W_1^2W_5=1& &W_1^2Z=1& & & \\
&W_1W_4^2=-1& &W_1W_5^2=-1& &W_1Z^2=-3& &Z^3=9,&\\ 
\end{aligned}
\ee
we obtain
\be
\begin{aligned}
Z_S=&-{\cal G}_1+{\cal G}_3+\frac{X^1}{2}+X^0,\cr
Z_{S'}=&-{\cal G}_2+\frac{X^2}{2}+X^0.
\end{aligned}
\ee
These expressions can be analytically continued to a neighborhood 
of $\CQ$ using again the prescription of \cite{Aspinwall:1994zu}.
We obtain in the leading order 
\be\label{eq:centralch}
\begin{aligned}
{\widetilde Z}_S=&{\widetilde X}^0-\frac{i\Gamma\left(-\frac{1}{6}\right)\Gamma\left(\frac{1}{6}\right)}{24\pi^{3/2}
\Gamma\left(\frac{2}{3}\right)\Gamma\left(\frac{5}{6}\right)}w_1+\cdots,\cr
{\widetilde Z}_{S'}=&{\widetilde X}^0-\frac{i\Gamma\left(-\frac{1}{6}\right)\Gamma\left(\frac{1}{6}\right)}{24\pi^{3/2}
\Gamma\left(\frac{2}{3}\right)\Gamma\left(\frac{5}{6}\right)}w_2+\cdots.\cr
\end{aligned}
\ee
Now we can conclude the proof of assertions $(a)$ and $(b)$ in section three. 
Expressions \eqref{eq:lognewper} show that indeed the quantum volumes of 
$\Sigma, \Sigma'$ vanish identically along $\CQ$. Moreover, according 
to equation \eqref{eq:centralch}, ${\widetilde Z}_S, {\widetilde Z}_{S'}$ 
are equal to ${\widetilde X}^0$ along $\CQ$ as claimed in section three. 
 
To conclude this section, note that one can similarly construct 
elliptic fibration with involution which admit $dP_7$ and $dP_6$ 
contractions respectively. These models can be obtained by replacing the 
ambient $\IP^2_{[1,2,3]}$-toric fibration employed in the above 
construction by $\IP^2_{[1,1,2]}$ and $\IP^2_{[1,1,1]}$ fibrations respectively, 
according to \cite{Klemm:1996hh}.

\subsection{A Quintic Model} 

Next we construct a different model involving quotients of quintic 
threefolds by holomorphic involutions. 

Let $Z$ be the blow-up of 
$\IP^4$ at two distinct points $p_1,p_2$. For concreteness we will 
choose homogeneous coordinates $[s,t,z^1,z^2,z^3]$ on $\IP^4$ so that 
the two points are given by 
\[
p_1= [ 1,0,0,0,0]\qquad 
p_2= [0,1, 0,0,0].
\] 
Let $E_1,E_2$ denote the exceptional divisors on $Z$; $E_1,E_2$ 
are isomorphic to $\IP^3$. 
The anticanonical class of $Z$ is $-K_Z =5H -3E_1-3E_2$, where 
$H$ denotes the pull-back of the hyperplane class on $\IP^4$ to 
$Z$. Note that if $X$ is a smooth divisor in the anticanonical 
linear system $|-K_Z|$, then $X$ is Calabi-Yau and contains 
two del Pezzo surfaces $S_1,S_2 \simeq dP_6$ obtained by 
intersecting $X$ with the two exceptional divisors $E_1,E_2$.
We will obtain a new model if 
we can also find an antiholomorphic involution of 
$X$ which maps $S_1$ isomorphically to $S_2$, having at the same time 
codimension $1$ or $3$ fixed loci.  
Therefore our goal is to show 
that the generic anticanonical divisor on $Z$ is smooth, and exhibit 
a family of such smooth divisors equipped with holomorphic involutions.

Let us first prove that the generic anticanonical divisor on $Z$ is 
smooth. Let $p:Z\to \IP^4$ denote the blow-up map, which contracts 
the exceptional divisors $E_1,E_2$. Given any anticanonical divisor 
$X$ on $Z$, the image of $X$ under $p$ is a quintic hypersurface 
$X'$ in $\IP^4$ with at least two cubic singularities at $p_1,p_2$. 
Conversely, the strict transform of any such quintic hypersurface
$X'$ is an anticanonical divisor on $X$. 
Therefore we have a one-to-one correspondence between anticanonical 
divisors $X$ on $Z$ and quintic hypersurfaces $X'$ with at least cubic 
singularities at $p_1,p_2$. 

Now let us analyze the sub-linear system of quintic hypersurfaces 
$X'$ in $\IP^4$. The generic hypersurface in this sub-linear system 
has an equation of the form 
\be\label{eq:quintB} 
\mathop{\sum_{m,n=0}^2\ \ \sum_{r_1,r_2,r_3=0}^5}_{m+n+r_1+r_2+r_3=5} 
a_{m,n,r_1,r_2,r_3}  s^mt^n(z^1)^{r_1}(z^2)^{r_2}(z^3)^{r_3}=0.
\ee
The base locus of this sub-linear system is the line 
$L\subset \IP^4$ cut by the 
equations 
\be\label{eq:baselocA}
z^1=z^2=z^3=0.
\ee
Indeed it is easy to check that given a point $p\in \IP^4\setminus L$, 
one can always find a quintic in the sub-linear system \eqref{eq:quintB} 
not passing through $p$. 

What we actually need is the base locus of the linear system of
hypersurfaces $X$ in $Z$. From the above it follows that this is
contained in the total transform of the line $L$, which equals the
proper transform of $L$ plus the two exceptional divisors $E_1,E_2$. 
But in fact,
the base locus is just the proper transform of $L$: for any point $q$ of
$E_i$, $i=1,2$, other than its intersection point $q_i$ with $L$, we can
find an $X$ not passing through $q$. This can be achieved, e.g., by taking
the quintic\footnote{Abusing notation, in the following
we will use the same notation for a hypersurface and its defining 
polynomial.} $X'=GA_1A_2$ to consist of a generic hyperplane
$G(z^1,z^2,z^3)=0$ through $L$ plus two generic quadratic cones $A_i$, with
vertices at the $p_i$, $i=1,2$. The corresponding $X$ meets 
$E_i$ in a generic plane through $q_i$ plus a generic
quadratic surface. The intersection of all such is clearly just the
point $q_i$.

As in the previous case, according to Bertini's theorem
a generic hypersurface can be singular
only at points of the base locus. 
In our case, this is the proper
transform of $L$. The $X$ corresponding to the above $X'=GA_1A_2$ is
singular at the two points $q_i$, $i=1,2$, but this is easy to remedy. We
consider a smooth cubic surface in the first exceptional divisor $E_1$, which
can be taken of the form:
\[
S_1=F(z^1, z^2, z^3) + t^2G(z^1, z^2, z^3),
\]
where $F$ is a smooth cubic in the z plane (e.g. the Fermat) and $G$ is a
general linear polynomial. Then the product $X_1=s^2S_1$ is in our
linear system and meets $E_1$ in the smooth cubic surface
$S_1$. This quintic is still singular at points (including $q_2$) of the
second exceptional divisor $E_2$, but the combination
\[X' := X_1 + X_2 = (s^2+t^2)F + 2s^2t^2G,\]
where $X_2$ is obtained from $X_1$ by interchanging $s$ and $t$, 
lifts to an $X$
which is smooth at all points of $L$. Bertini's theorem now tells us
that a generic $X$ in our linear system is smooth everywhere.

Having gone this far, we can actually write down an explicit smooth $X$,
thus avoiding the need to use Bertini. 
Let us consider the following family of hypersurfaces of the form 
\eqref{eq:quintB}
\be\label{eq:quintC} 
P\equiv (s^2+t^2)F(z^1,z^2,z^3) + s^2t^2G(z^1,z^2,z^3) + Q(z^1,z^2,z^3)=0
\ee 
where 
\[
\begin{aligned} 
F(z^1,z^2,z^3) & = \sum_{i=1}^3 a_i(z^i)^3 \\
G(z^1,z^2,z^3) & = \sum_{i=1}^3 b_i z^i \\
Q(z^1,z^2,z^3) & = \sum_{i=1}^3 c_i z_i^5\\
\end{aligned}
\]
and $(a_i,b_i,c_i)$, $i=1,2,3$ are generic coefficients. 
Again, one can check that the base locus of this family is 
$L$, therefore the generic hypersurface of the form 
\eqref{eq:quintC} may be singular only at points on $L$. 
Note that 
\[
\partial_{z^i}P(s,t,0,0,0) = s^2t^2b_i, \quad i=1,2,3
\]
Therefore if at least one of the $b_i$, $i=1,2,3$
is nonzero, the singularities of $P$ are located at 
$s=0$ or $t=0$. Therefore for generic $b_i$, $i=1,2,3$ 
the hypersurfaces \eqref{eq:quintC} have singularities only 
at $p_1,p_2$. 

Next, let us check that the exceptional divisors 
$S_1,S_2$ on the strict transform $X$ of a generic 
hypersurface of the form \eqref{eq:quintC} are smooth. 
It suffices to prove this statement for only one 
of the exceptional divisors, say $S_1$, since the computations 
are identical. We will work in the affine coordinate chart 
$s\neq 0$. Abusing notation we will denote the affine coordinates 
in this chart by $(t,z^1,z^2,z^3)$. Then the equations
\eqref{eq:quintC} become 
\be\label{eq:quintD} 
(1+t^2)F(z^1,z^2,z^3) + t^2 G(z^1,z^2,z^3) + Q(z^1,z^2,z^3)=0.
\ee
The blow-up $Z$ is locally isomorphic to the one point blow-up 
of the affine chart $s\neq 0$ at $p_1$. Let $[\rho,\lambda^1,
\lambda^2,\lambda^3]$ denote homogeneous coordinates on $\IP^3$. 
Then $Z$ is locally described by the following equations 
\[
\begin{aligned}
t\lambda^i & = \rho z^i , \qquad i=1,2,3\\
z^i\lambda^j& = z^j \lambda^i,\qquad i,j=1,2,3,\ i\neq j \\
\end{aligned}
\]
in $\IC^4\times \IP^3$. 
Taking into account \eqref{eq:quintD}, the 
exceptional divisor $S_1$ is the cubic hypersurface
\be\label{eq:cubicA}
\sum_{i=1}^3 a_i (\lambda^i)^3 + \rho^2 \sum_{i=1}^3 b_i \lambda^i 
=0.
\ee
The singular points of the cubic hypersurface \eqref{eq:cubicA} 
are determined by 
\be\label{eq:cubicB}
\begin{aligned} 
3a_i (\lambda^i)^2 + \rho^2 b_i & =0, \qquad i=1,2,3\\
\end{aligned}
\ee
in addition to equation \eqref{eq:cubicA}. 
Assuming $a_i\neq 0$, for $i=1,2,3$, let $\omega_i$ be a 
square root of $-b_i/3a_i$, $i=1,2,3$. Then the solutions 
of the equations \eqref{eq:cubicB} are of the form  
$\lambda^i = \omega_i \rho$, and $S_1$ is singular if and only if 
the following relation holds 
\be\label{eq:cubicC} 
\begin{aligned}
\sum_{i=1}b_i\omega_i& =0\cr
\end{aligned}
\ee
Equation \eqref{eq:cubicC} is not satisfied for 
generic values of $(a_i,b_i)$, 
$i=1,2,3$, therefore the generic cubic is smooth.

Next, we have to exhibit a subfamily of smooth anticanonical 
divisors $X$ on $Z$ equipped with holomorphic involutions 
$\sigma:X\to X$ which exchange $S_1,S_2$ and have codimension 
$1$ or $3$ fixed loci. Note that it suffices to construct
hypersurfaces $X'$ in the sub-linear system \eqref{eq:quintB} 
equipped with holomorphic involutions $\sigma':X'\to X'$ so that 
$\sigma'(p_1)=p_2$ and the fixed locus $(X')^{\sigma'}$ has 
components of dimension zero or two supported away from $p_1,p_2$. 
Any such involution lifts to an involution $\sigma:X \to X$ 
on the strict transform of $X'$ with the desired properties. 

Consider the holomorphic involution 
\[
\tau: \IP^4\to \IP^4, \qquad 
\tau[s,t,z^1,z^2,z^3] = [t,s,-z^1,-z^2,-z^3].
\]
Note that $\tau$ preserves 
equation \eqref{eq:quintC}, therefore it induces a 
holomorphic involution $\sigma': X'\to X'$
on each hypersurface $X'$ in the family \eqref{eq:quintC}. 
The fixed locus of $\sigma'$ consists of the point 
$p=[1,1,0,0,0]$, 
and the divisor cut by the equation $s+t=0$. 
Moreover, $\sigma'$ obviously exchanges $p_1,p_2$. 
Since $\tau$ preserves equation \eqref{eq:quintC} for 
arbitrary values of $(a_i,b_i,c_i)$, it follows that 
the strict transform $X$ is a smooth threefold on $Z$ equipped 
with a holomorphic involution $\sigma: X\to X$ which 
maps $S_1$ isomorphically to $S_2$. 

In order to conclude our construction, let us show that the 
generic hypersurface $X$ equipped with a holomorphic involution 
$\sigma$ satisfies conditions $(i)-(iv)$ formulated in section three. 
Condition $(i)$ is clearly satisfied. Condition two follows from the 
fact that the dominant monomial $stz^1z^2z^3$ in the large complex 
structure limit is odd under the involution $\tau$. Since the 
right hand side of equation \eqref{eq:quintD} is also odd, it follows 
that we can perturb the family \eqref{eq:quintD} by a term 
of the form $\mu stz^1z^2z^3$, with arbitrary $\mu$. The resulting 
hypersurfaces are still smooth if the coefficients $(a_i,b_i,c_i)$ are 
generic. Conditions $(iii)$ and $(iv)$ follow from the fact that 
$H^{1,1}(X)$ is generated by the divisor classes $S_1,S_2,H$, 
while $H^{2,2}(X)$ is generated by the curve classes $L,\Sigma_1,\Sigma_2$,
where $\Sigma_i = (S_i)^2_X$.

\appendix
\bigskip
\section{Complements on periods} 

\noindent For completeness we provide explicit expressions for the 
functions $h_i(z_1,\ldots,z_4)$ in the expressions \eqref{eq:logper}
of the periods $X^1,X^2$.
\be
\begin{aligned}
h_0=&\frac{z_1^{n_1}z_2^{n_2}z_3^{n_3}z_4^{n_4}}
{\Gamma(1+n_1-n_3+n_4)\Gamma(1+n_2+n_4)\Gamma(1+n_1+n_2-n_3+n_4)
\Gamma(1-n_1+n_3)\Gamma(1-n_2)}\cr
&\times\frac{\Gamma(1+6n_1)\Gamma(1+6n_2)}{\Gamma(1+n_3-3n_4)
\Gamma(1+2n_1+2n_2)
\Gamma(1+3n_1+3n_2)},\cr
h_1=&\frac{(-z_1)^{n_1}z_2^{n_2}(-z_3)^{n_3}(-z_4)^{n_4}}
{\Gamma(1+n_2+n_4)\Gamma(1+n_1+n_2-n_3+n_4)\Gamma(1-n_1+n_3)
\Gamma(1-n_2)\Gamma(1+n_3-3n_4)}\cr
&\times\frac{\Gamma(-n_1+n_3-n_4)\Gamma(1+6n_1)\Gamma(1+6n_2)}
{\Gamma(1+2n_1+2n_2)
\Gamma(1+3n_1+3n_2)},\cr
\end{aligned}
\ee
\[
\begin{aligned}
h_2=&\frac{(-z_1)^{n_1}(-z_2)^{n_2}(-z_3)^{n_3}(-z_4)^{n_4}}
{\Gamma(1+n_1-n_3+n_4)\Gamma(1+n_2+n_4)\Gamma(1-n_1+n_3)\Gamma(1-n_2)
\Gamma(1+n_3-3n_4)}\cr
&\times\frac{\Gamma(-n_1-n_2+n_3-n_4)\Gamma(1+6n_1)\Gamma(1+6n_2)}{\Gamma(1+2n_1+2n_2)
\Gamma(1+3n_1+3n_2)},\cr
h_3=&\frac{(-z_1)^{n_1}z_2^{n_2}(-z_3)^{n_3}z_4^{n_4}}
{\Gamma(1+n_1-n_3+n_4)\Gamma(1+n_2+n_4)\Gamma(1+n_1+n_2-n_3+n_4)\Gamma(1-n_2)\Gamma(1+n_3-3n_4)}\cr
&\times\frac{\Gamma(n_1-n_3)\Gamma(1+6n_1)\Gamma(1+6n_2)}{\Gamma(1+2n_1+2n_2)
\Gamma(1+3n_1+3n_2)},\cr
h_4=&\frac{z_1^{n_1}(-z_2)^{n_2}z_3^{n_3}z_4^{n_4}}
{\Gamma(1+n_1-n_3+n_4)\Gamma(1+n_2+n_4)\Gamma(1+n_1+n_2-n_3+n_4)\Gamma(1-n_1+n_3)}\cr
&\times\frac{\Gamma(n_2)\Gamma(1+6n_1)\Gamma(1+6n_2)}{\Gamma(1+n_3-3n_4)\Gamma(1+2n_1+2n_2)
\Gamma(1+3n_1+3n_2)}.
\end{aligned}
\]

\bibliography{qcd.bib}
\bibliographystyle{utcaps}

\end{document}